\documentclass[preprint,aps,prd,groupedaddress,showpacs,nofootinbib]{revtex4-1}%
\usepackage{graphicx}
\usepackage{amsmath}
\usepackage{amssymb}
\usepackage{bm}
\usepackage{epsfig}
\usepackage{dcolumn}
\usepackage{slashed}
\usepackage{color}

\def\LO{\text{LO}}
\def\NLO{\text{NLO}}
\def\NNLO{\text{NNLO}}

\begin{document}

\def\Maryland{Maryland Center for Fundamental Physics, University of Maryland, College Park, Maryland 20742, USA}
\def\Argonne{High Energy Physics Division, Argonne National Laboratory, Argonne, IL 60439, USA}
\def\Northwestern{Department of Physics \& Astronomy, Northwestern University, Evanston, IL 60208, USA}

\title{Phenomenology of the $Z$-boson plus jet process at NNLO}

\author{Radja Boughezal}
\email{rboughezal@anl.gov}
\affiliation{\Argonne}

\author{Xiaohui Liu}
\email{xhliu@umd.edu}
\affiliation{\Maryland}

\author{Frank Petriello}
\email{f-petriello@northwestern.edu}
\affiliation{\Argonne}
\affiliation{\Northwestern}

\begin{abstract}

We present a detailed phenomenological study of $Z$-boson production in association with a jet through next-to-next-to-leading order (NNLO) in perturbative QCD.  Fiducial cross sections and differential distributions for both 8 TeV and 13 TeV LHC collisions are presented.  We study the impact of different parton distribution functions (PDFs) on predictions for the $Z$+jet process.  Upon inclusion of the NNLO corrections, the residual scale uncertainty is reduced such that both the total rate and the transverse momentum distributions can be used to discriminate between various PDF sets.

\end{abstract}

\maketitle

\section{Introduction}

$Z$-boson production in association with a jet plays an important role in the physics program of the Large Hadron Collider (LHC).  A precise understanding of this process is critical in searches for physics beyond the Standard Model, since it is an important background to signatures of both dark matter and supersymmetric particle production.  $Z$+jet production  is a background to Higgs production, and an accurate description of the process is necessary to pursue the program of Higgs coupling measurements during Run II of the LHC.  In addition to these physics motivations, $Z$+jet production also serves as an important experimental benchmark.  It is needed for calibration of the jet energy scale, and there is hope that it will lead to an improved extraction of the gluon distribution function.

Motivated by its central role in the LHC program, numerous theoretical effort has been invested in precisely predicting this process.  The next-to-leading order (NLO) electroweak corrections were recently considered in Ref.~\cite{EWcorZ}.  A merged NLO QCD+electroweak prediction was obtained in Ref.~\cite{NLOQCDEW}.  The leading threshold logarithms beyond NLO in QCD have been considered~\cite{Becher:2011fc}.  Very recently the full next-to-next-to-leading order (NNLO) QCD predictions have been derived using two different theoretical techniques~\cite{Ridder:2015dxa,Boughezal:2015ded}.   In recent work we have shown that the NNLO QCD corrections are essential in achieving an accurate description of several 7 TeV LHC results~\cite{Boughezal:2016yfp}.  The NNLO predictions promise to make possible similar precision comparisons of theory with LHC Run II data. 

It is our goal in this manuscript to present a detailed study of the NNLO QCD predictions for $Z$+jet production for both 8 TeV and 13 TeV LHC collisions.  We present predictions for the leading jet and $Z$-boson differential distributions over the entire accessible kinematic range.  We also study the distributions of the leptons that arise from the $Z$-boson decay.  It has been pointed out in the literature that the $Z$-boson transverse momentum distributions could provide an important constraint on the high-$x$ gluon distribution function~\cite{Malik:2013kba}.  We therefore study the dependence of the fiducial cross section and distribution shapes on the choice of parton distribution function (PDF).  The major findings of our paper are summarized below.
\begin{itemize}

\item The NNLO QCD corrections to the fiducial cross sections and to most distributions are at the few percent level.  The residual theoretical uncertainties are reduced to the percent level at NNLO.

\item The corrections to the high transverse momentum region of the leading jet and to the high $H_T$ region, where $H_T$ is the scalar sum of the jet transverse momenta, are large.  This is similar to the behavior seen for the $W$+jet process~\cite{Boughezal:2016dtm}.  As discussed in the text, we do not expect further large N$^3$LO corrections to this distribution.

\item Both the fiducial cross section and the shapes of the transverse momentum distributions are sensitive to the choice of PDF.  The differences between the sets can be larger than the estimated theoretical uncertainty, suggesting that the $Z$+jet process will eventually provide a useful constraint on PDF fits.

\end{itemize}

Our paper is organized as follows.  We discuss our calculational setup and all parameter choices in Section~\ref{sec:setup}.  Numerical results for the fiducial cross sections and differential distributions for 8 TeV and 13 TeV collisions are presented in Section~\ref{sec:numgen}.  A comparison of the predictions from different PDF sets is performed in Section~\ref{sec:PDFs}.  Finally, we conclude in Section~\ref{sec:conc}.

\section{Setup}
\label{sec:setup}

We discuss here our calculational setup for $Z$-boson production in association with a jet through NNLO in perturbative QCD.  We study collisions in the inclusive one-jet bin for both 8 TeV and 13 TeV LHC energies.  Jets are defined using the anti-$k_t$ algorithm~\cite{Cacciari:2008gp} with $R=0.4$.  We study several different parton distribution function extractions: CT14~\cite{Dulat:2015mca}, NNPDF3.0~\cite{Ball:2014uwa}, MMHT2014~\cite{Harland-Lang:2014zoa}, and ABM12~\cite{Alekhin:2013nda}.  When computing a hadronic cross section we use them at the appropriate order in perturbation theory: LO PDFs together with a LO partonic cross section, NLO PDFs with a NLO partonic cross section, and NNLO PDFs with a NNLO partonic cross section.  We choose the central scale 
\begin{equation}
\mu_0 = \sqrt{M_{ll}^2+\sum_i (p_T^{J_i})^2}
\end{equation}
for both the renormalization and factorization scales, where $M_{ll}$ is the invariant mass of the dilepton system and the sum $i$ runs over all reconstructed jets.  This dynamical scale correctly captures the characteristic energy throughout the entire kinematic range studied here, which extends into the TeV region.  To estimate the theoretical uncertainty we vary the renormalization and factorization scales independently in the range $\mu_0/2 \leq \mu_{R,F} \leq 2 \mu_0$, subject to the restriction
\begin{equation}
1/2 \leq \mu_R / \mu_F \leq 2.  
\end{equation}
All presented results assume decay into a single massless lepton species, and include contributions from both the $Z$-boson and an off-shell photon.

We consider in this paper a selection criterion that matches what ATLAS has used in their studies of the $Z$+jet process at 13 TeV\footnote{We thank U.~Blumenschein and J.~Huston regarding the selection cuts used by ATLAS}.  These are similar to the cuts imposed in the 8 TeV ATLAS and CMS analyses~\cite{Aad:2013ysa,Khachatryan:2014zya}, and follow what we have used in our previous study of $W$-boson plus jet production~\cite{Boughezal:2016dtm}.  We impose the following fiducial cuts:
\begin{equation}
\begin{split}
p_T^{J} &> 30 \, \text{GeV}, \;\;\; |\eta^J|<2.5,\;\;\; p_T^{J_1}>100 \, \text{GeV}, \\
p_T^{l} &> 25 \, \text{GeV}, \;\;\; |\eta^{l}| < 2.5, \;\;\; 71 \, \text{GeV} < M_{ll}<  111 \, \text{GeV}.
\end{split}
\label{eq:cuts1}
\end{equation}
$p_T^{J_1}$ refers to the transverse momentum of the leading jet, on which we have imposed an additional cut, while $\eta^J$ refers to the jet pseudorapidity.  $p_T^l$ and $\eta^l$ refer to the transverse momenta and pseudorapidities of the leptons.  We study the following distributions: $p_T^{J_1}$, $p_T^Z$, $\eta^{J_1}$, $Y_Z$, $H_T$, $p_T^{l,h}$, $p_T^{l,s}$.  Here, $Y_Z$ denotes the rapidity of the dilepton pair, $\eta^{J_1}$ the pseudorapidity of the leading jet, and $p_T^Z$ the transverse momentum of the dilepton pair.  $H_T$ is the scalar sum of the transverse momenta of all reconstructed jets.  The superscripts $h$ and $s$ for the lepton distributions refer to the hardest and softest leptons, respectively.  All of these distributions begin first at leading order for the $Z$+jet process.

The NNLO calculation upon which our phenomenological study is based was obtained using the $N$-jettiness subtraction scheme~\cite{Boughezal:2015dva,Gaunt:2015pea}.  This method relies upon splitting the real-emission phase space according to the $N$-jettiness variable $\tau_N$~\cite{Stewart:2010tn}.  For values of $N$-jettiness greater than some cut, $\tau_N > \tau_N^{cut}$, an NLO calculation for $Z$+2-jets is used.  Any existing NLO program can be used to obtain these results.  We use the public code MCFM~\cite{Campbell:2010ff,Campbell:2015qma} in this study.  For the phase-space region $\tau_N < \tau_N^{cut}$, an all-orders resummation formula is used to obtain the contribution to the cross section~\cite{Stewart:2010tn,Stewart:2009yx,Gaunt:2014xga,Becher:2006qw,Boughezal:2015eha}.  This formulation relies heavily upon the theoretical machinery of soft-collinear effective theory~\cite{scet}.  An important check of this formalism is the independence of the full result from $\tau_N^{cut}$.  The application and validation of $N$-jettiness subtraction for one-jet processes has been discussed several times in the literature, including for the $Z$+jet process~\cite{Boughezal:2015dva,Boughezal:2015aha,Boughezal:2015ded}.  We do not review this topic here.  We have computed each bin of the studied distributions for several $\tau_N^{cut}$ values, and have found independence of our results from $\tau_N^{cut}$ within numerical errors.
 
\section{Numerical results}
\label{sec:numgen}

We begin by discussing the fiducial cross sections for both 8 TeV and 13 TeV collisions using the cuts in Eq.~(\ref{eq:cuts1}).  For all cross sections in this section we use CT14 PDFs.  Results for other PDF choices are given in the next section. The LO, NLO, and NNLO 1-jet cross sections, as well as the $K$-factors $K_{\NLO} = \sigma_{\NLO}/\sigma_{\LO}$ and $K_{\NNLO} = \sigma_{\NNLO}/\sigma_{\NLO}$, are presented in Table~\ref{tab:fidinc}.  For both energies there is an approximately 60\% increase of the cross section in going from LO to NLO, with a slightly larger correction occurring for $\sqrt{s}=13$ TeV.  The NNLO corrections are smaller, and increase the NLO result by only 4\% for the central scale choice.  This indicates the good convergence of QCD perturbation theory for the fiducial cross section.  The residual errors as estimated by scale variation decrease from the approximately 10\% level at NLO to the percent level at NNLO.
 
\begin{table}[h]
\begin{tabular}{|c|c|c|c|c|c|}
\hline
 & $\sigma_{\LO}$ (pb) & $\sigma_{\NLO}$ (pb) & $\sigma_{\NNLO}$ (pb) & $K_{\NLO}$ & $K_{\NNLO}$ \\
 \hline\hline
8 TeV & $4.01^{+0.52}_{-0.43}$ & $6.39^{+0.60}_{-0.51}$ & $6.72^{+0.01}_{-0.15}$ & 1.59 & 1.05 \\
13 TeV & $8.53^{+0.80}_{-0.73}$ & $14.12^{+1.23}_{-1.00}$ & $14.87^{+0.05}_{-0.28}$ & 1.65 & 1.05 \\
\hline
\end{tabular}
\caption{Fiducial cross sections for the $Z$+jet process for 8 TeV and 13 TeV collisions, using the cuts of Eq.~(\ref{eq:cuts1}).  The scale errors are shown for the LO, NLO and NNLO cross sections.  The $K$-factors are shown for the central scale choice.}
\label{tab:fidinc}
\end{table} 

\begin{figure}[h]
    \includegraphics[width=.49\linewidth]{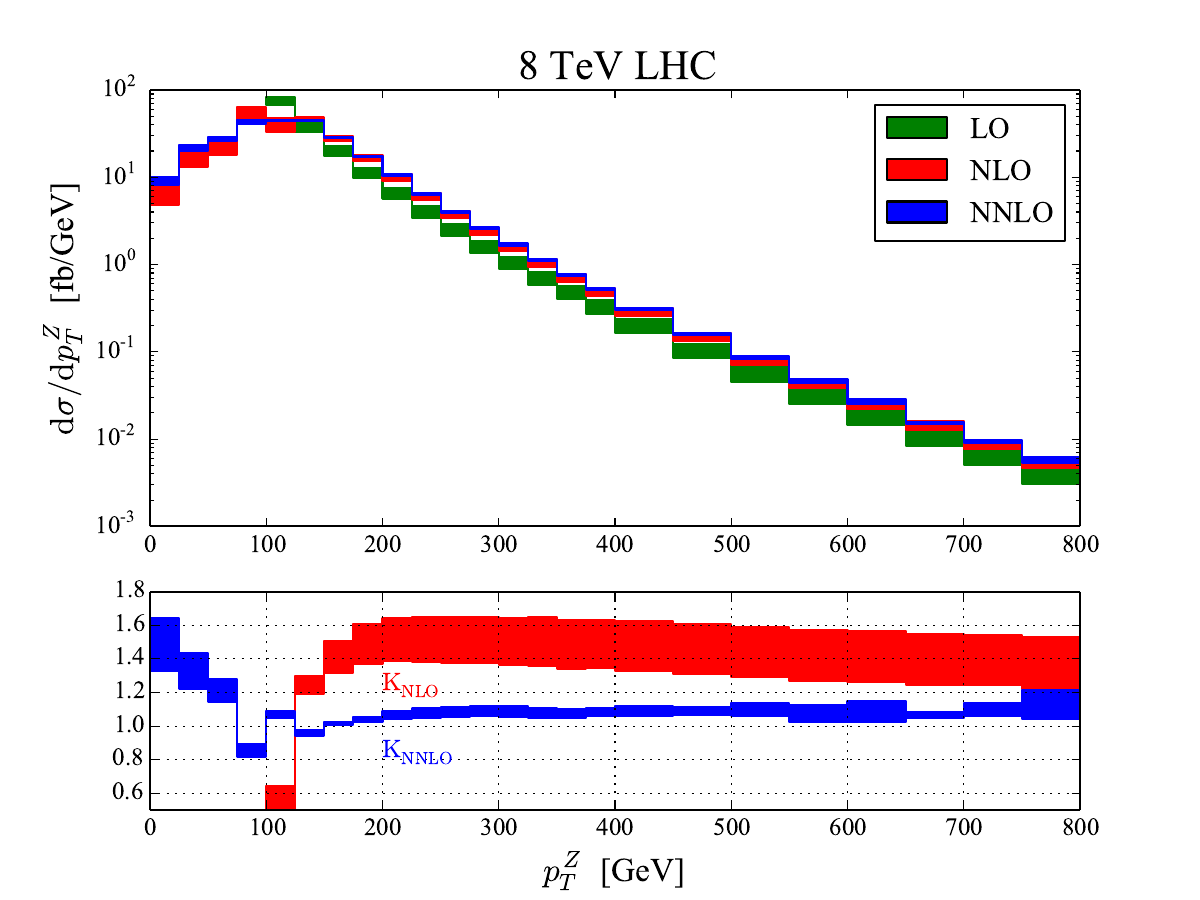} 
    \includegraphics[width=.49\linewidth]{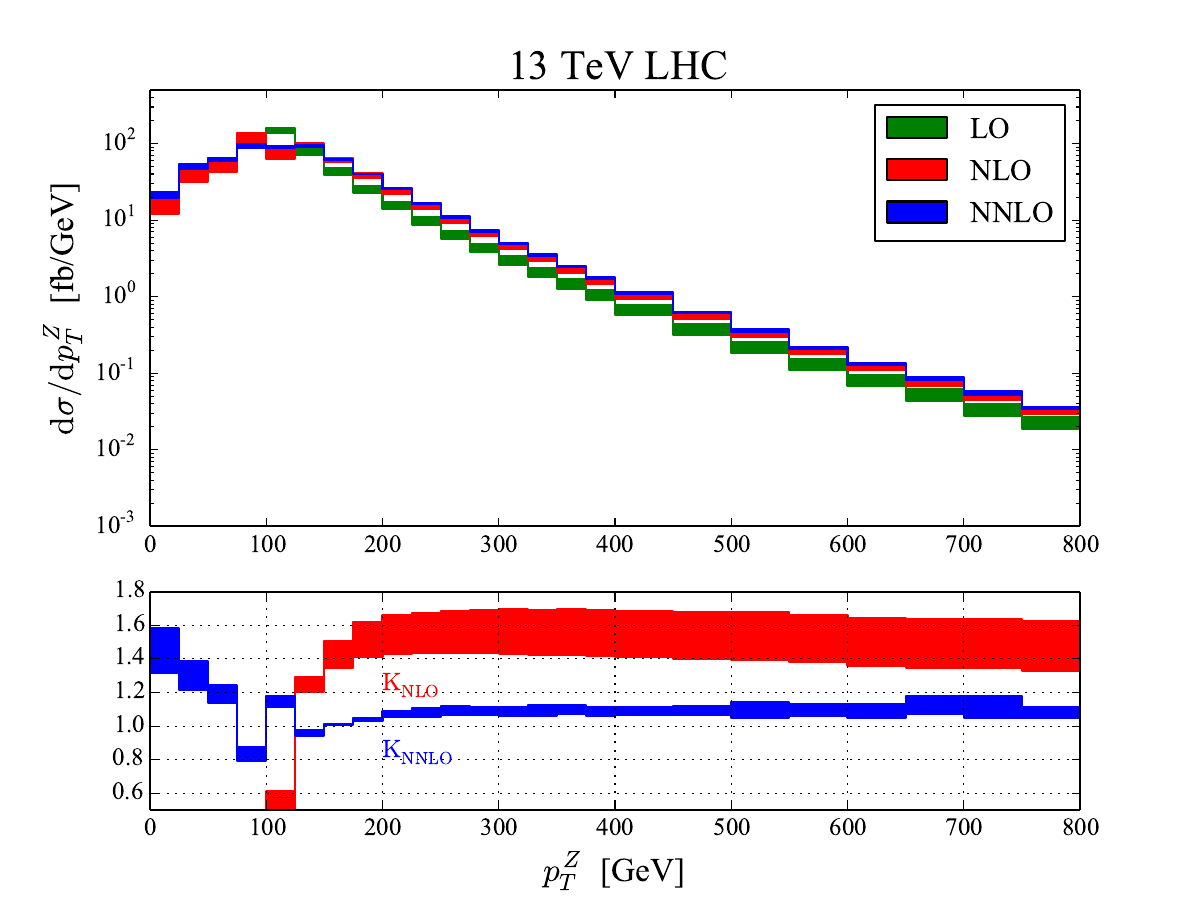} 
    \caption{Plots of the $Z$-boson transverse momentum distribution for both 8 TeV and 13 TeV collisions.  In each plot the upper inset shows the LO, NLO and NNLO distributions, while the lower inset shows $K_{\NLO}$ and $K_{\NNLO}$.  The bands indicate the scale variation.}
    \label{fig:pTZ}
\end{figure}

We now start our study of differential distributions in the $Z$+jet process with the transverse momentum distribution of the $Z$-boson.  The results for both 8 TeV and 13 TeV collisions are shown in Fig.~\ref{fig:pTZ}.  We note that the distribution is studied over a larger range than in previous work~\cite{Boughezal:2015ded}.  The NLO corrections decrease from a maximum of 60\% for $p_T^Z$ in the range 200-300 GeV to 40\% for $p_T^Z \approx 1$ TeV.  The NNLO corrections increase as the transverse momentum of the $Z$-boson is increased, rising to a maximum of 15\% at $p_T^Z \approx 1$ TeV.  The $K$-factors for 8 TeV and 13 TeV collisions exhibit similar dependence on $p_T^Z$, as shown in the lower inset.  In these plots and in all plots in this manuscript, the cross sections in the denominators of the $K$-factors are computed for their central scale choice.  The scales are varied in the numerator cross sections of the $K$-factors, leading to the bands shown.Upon inclusion of the NNLO corrections the scale dependence decreases to the $\pm 2-3\%$ level.  The behavior of the cross section near $p_T^Z=100$ GeV has been observed for the $W$+jet process as well~\cite{Boughezal:2016dtm}.  The leading-jet transverse momentum restriction $p_T^{J_1}> 100$ GeV implies that at LO, $p_{TZ}>100$ GeV.  This restriction is relaxed at NLO.  Near this kinematic boundary the cross section is sensitive to soft-gluon radiation, leading to the large corrections seen in Fig.~\ref{fig:pTZ}.

\begin{figure}[h]
    \includegraphics[width=.49\linewidth]{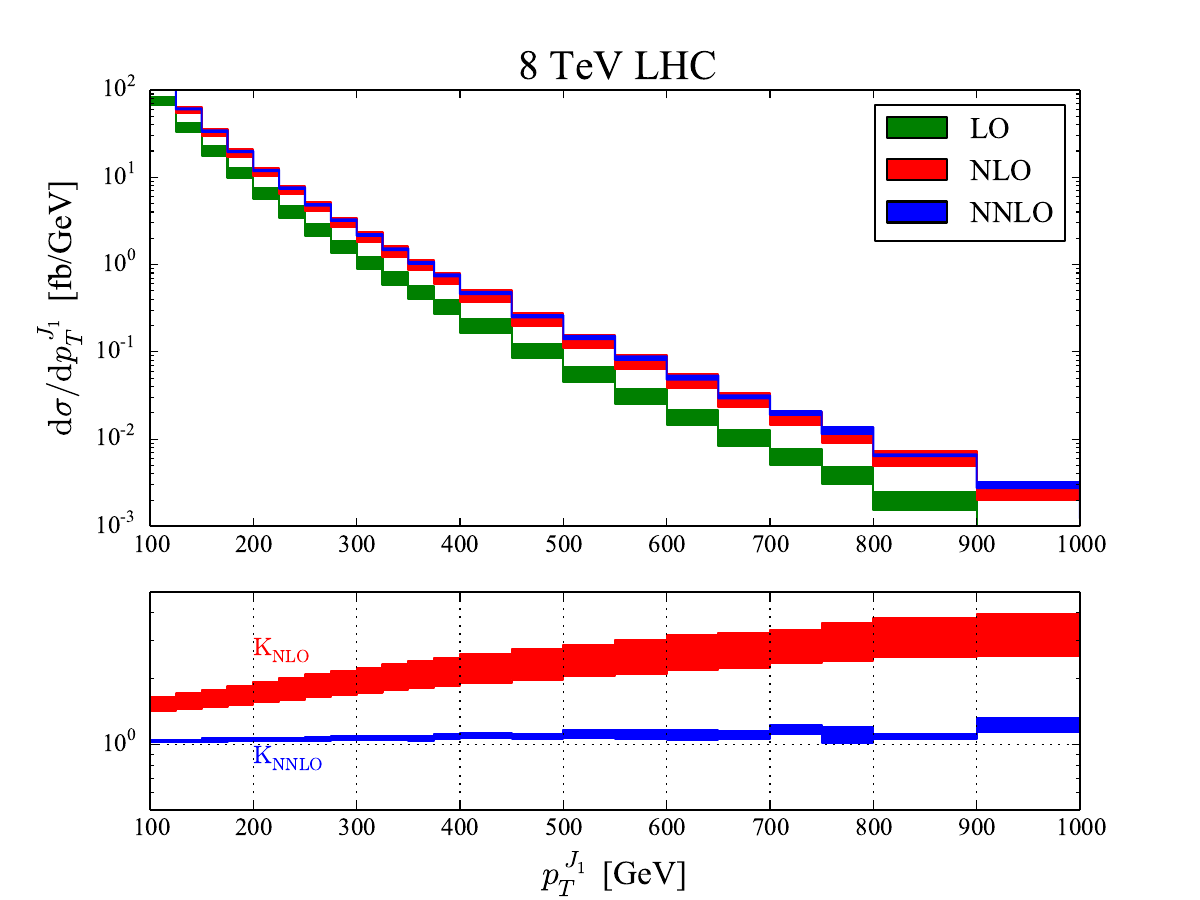} 
    \includegraphics[width=.49\linewidth]{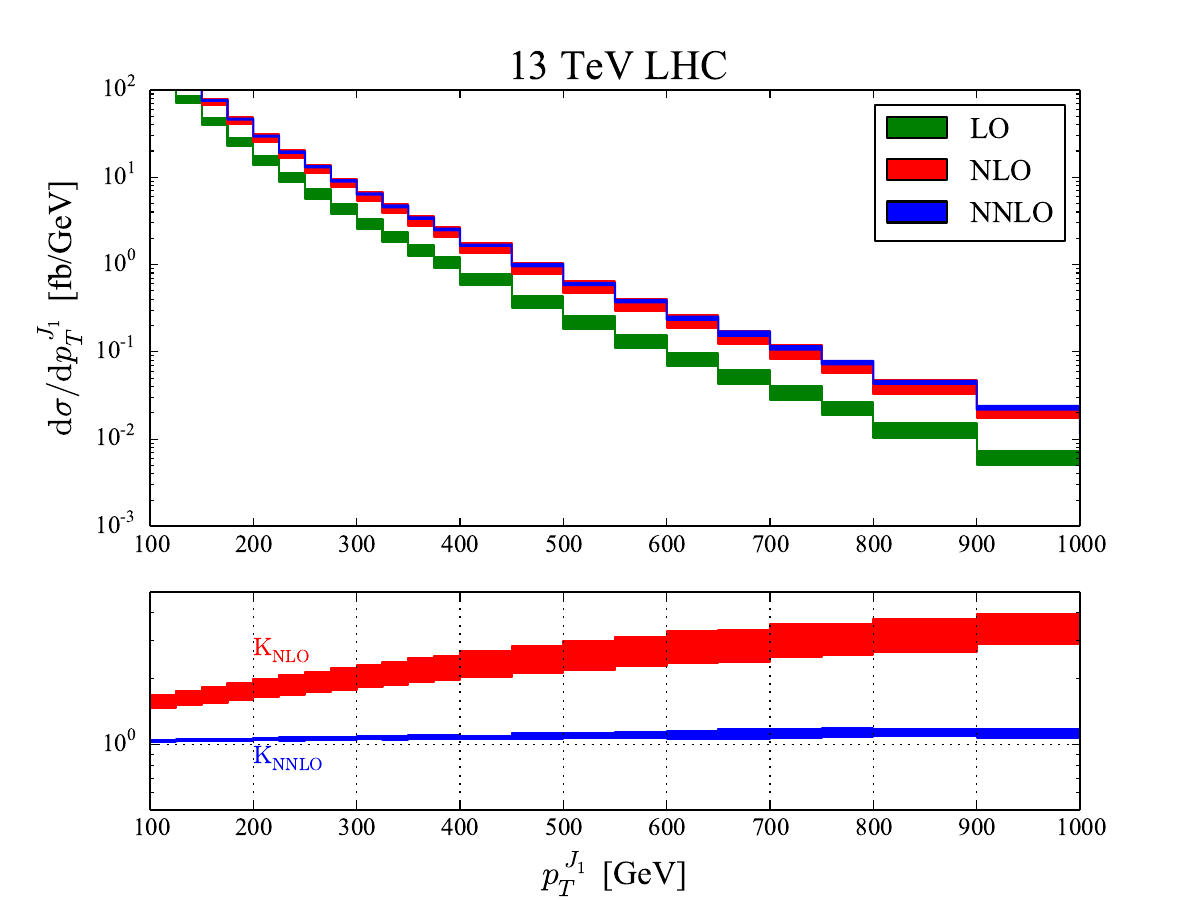} 
    \caption{Plots of the transverse momentum distribution of the leading jet for both 8 TeV and 13 TeV collisions.  In each plot the upper inset shows the LO, NLO and NNLO distributions, while the lower inset shows $K_{\NLO}$ and $K_{\NNLO}$.  The bands indicate the scale variation.}
    \label{fig:pTJ1}
\end{figure}

We next consider the transverse momentum distribution of the leading jet in Fig.~\ref{fig:pTJ1}.  The NLO $K$-factor grows to over a factor of 3 for $p_T^{J_1}>1$ TeV.  The reason for this behavior in $V$+jet processes is well-known~\cite{Rubin:2010xp}.   At NLO there exist configurations containing two hard jets and a soft or collinear $Z$ boson that are logarithmically enhanced. These cannot occur at LO, since the $Z$-boson must balance in the transverse plane against the single jet that appears.  Their appearance at NLO represents a qualitatively new effect that leads to the enhancement.  The inclusion of the NNLO terms stabilizes the perturbative expansion, with the NNLO $K$-factor reaching a maximum of 1.2 for $p_T^{J_1} > 1$ TeV.  The residual scale dependence at NNLO grows slightly as $p_T^{J_1}$ is increased, growing from 1-2\% for $p_T^{J_1}<400$ GeV to 10\% for $p_T^{J_1}>1$ TeV.
 
\begin{figure}[h]
    \includegraphics[width=.49\linewidth]{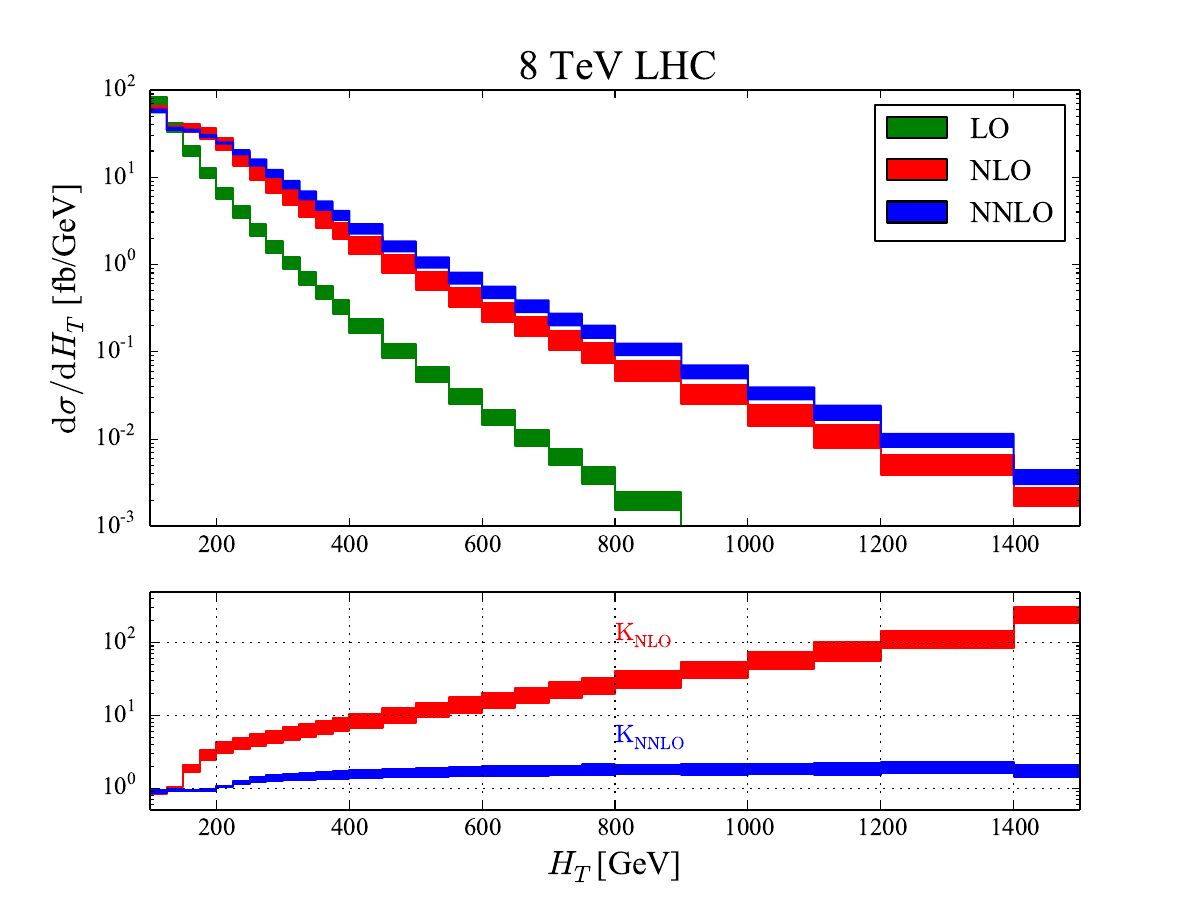} 
    \includegraphics[width=.49\linewidth]{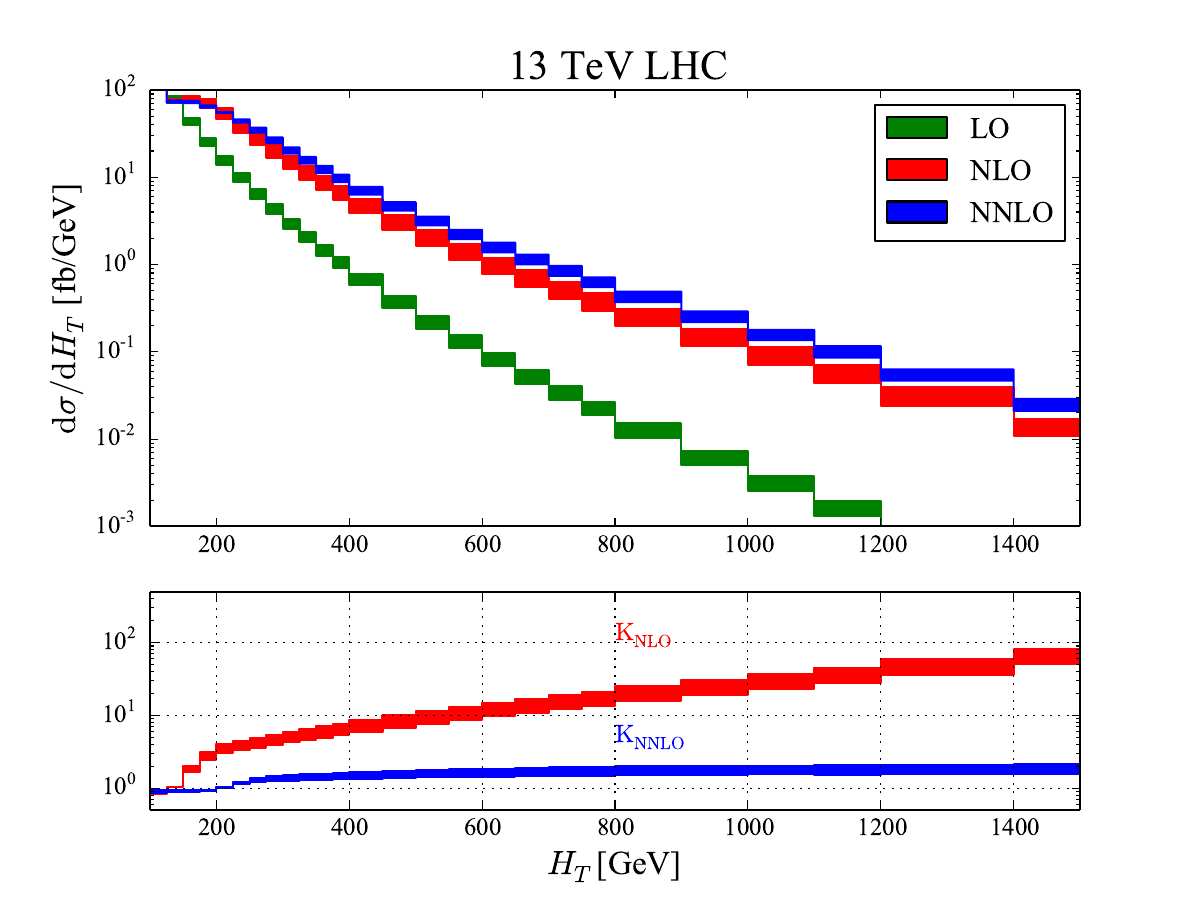} 
    \caption{Plots of the $H_T$ distribution for both 8 TeV and 13 TeV collisions.  In each plot the upper inset shows the LO, NLO and NNLO distributions, while the lower inset shows $K_{\NLO}$ and $K_{\NNLO}$.  The bands indicate the scale variation.}
    \label{fig:HT}
\end{figure}

The $H_T$ distribution, where $H_T$ is defined as the scalar sum of the transverse momenta of all reconstructed jets, is shown in Fig.~\ref{fig:HT}.  The increase of the cross section with $H_T$ is more dramatic than for the $p_T^{J_1}$ distribution.  It increases by more than a factor of 100 for $p_T^{J_1}>1.2$ TeV for 8 TeV collisions, and above a factor of 100 for $p_T^{J_1}>2$ TeV in 13 TeV collisions.  The explanation is the same as for the $p_T^{J_1}$ observables, except that the logarithmic enhancement is larger for this distribution~\cite{Rubin:2010xp}.  The NNLO corrections are still large, reaching a factor of two for $p_T^{J_1} \approx 1.5$ TeV in 8 TeV collisions, and for $p_T^{J_1} \approx 2$ TeV in 13 TeV collisions.  Such a large increase is needed to explain the available 7 TeV data from ATLAS and CMS~\cite{Boughezal:2016yfp}.  The residual scale dependence is still large at NNLO.  It grows to over $\pm 15\%$ for $H_T > 1.5$ TeV for both collision energies.  We note that the NNLO and NLO scale variation bands do not overlap at intermediate-to-large $H_T$.  Due to the dominance of the two-jet configuration at large $H_T$, the order of perturbation theory is effectively reduced in this region, rendering the NLO calculation effectively LO, and the NNLO result effectively NLO.  It is therefore not surprising that an effectively LO scale variation fails to properly estimate the size of higher-order corrections.
 
The large corrections to the $H_T$ distribution from two-jet processes begs the question of whether even higher-multiplicity processes may result in large N$^3$LO corrections to this observable.  It has been found for the similar $W$+jet process that merged multi-jet plus parton-shower simulations agree well with our NNLO result for the $H_T$ distribution~\cite{CMS-PAS-SMP-14-023}.  These predictions contain up to four hard jets, and therefore probe the impact of higher-multiplicity processes on our NNLO result.  The observed agreement indicates that further large shifts in this observable are unlikely at the N$^3$LO order in QCD.

\begin{figure}[h]
    \includegraphics[width=.49\linewidth]{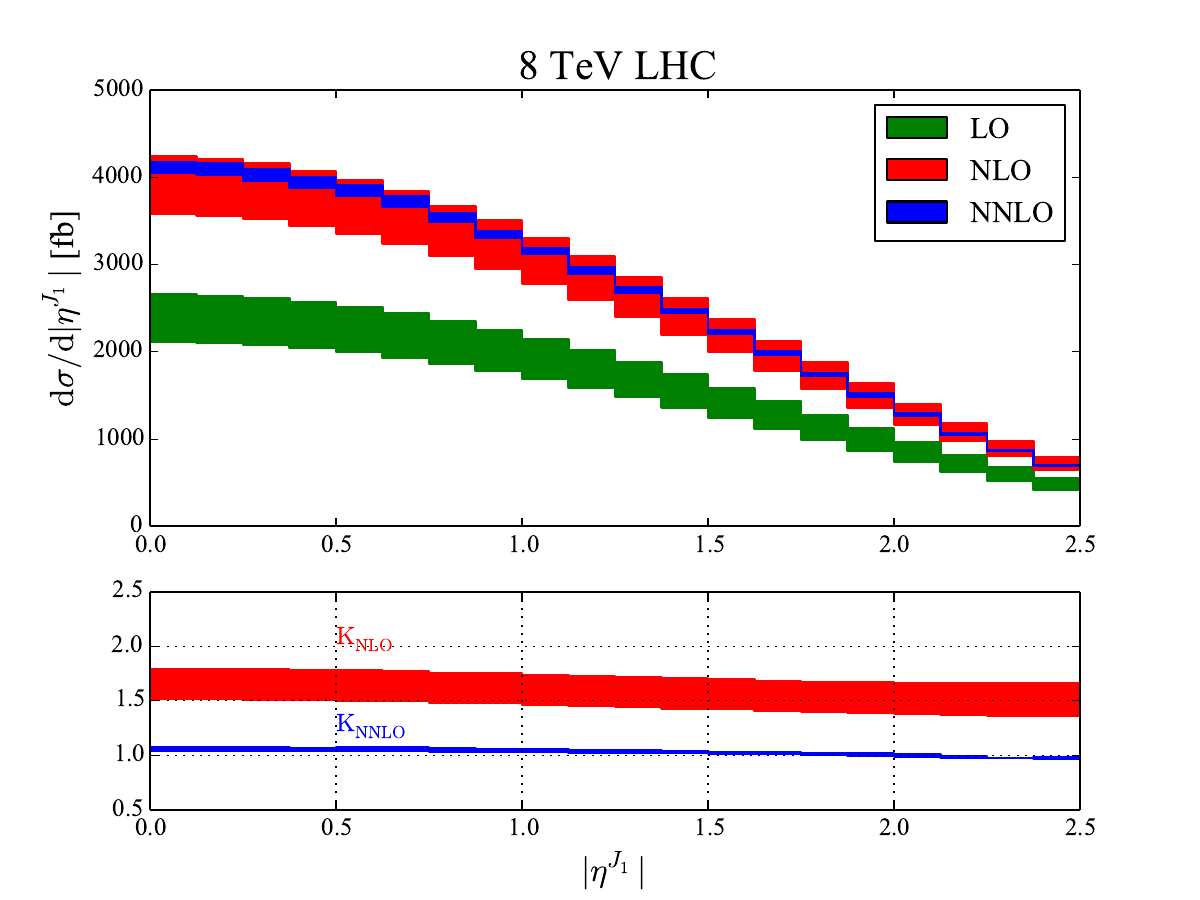} 
    \includegraphics[width=.49\linewidth]{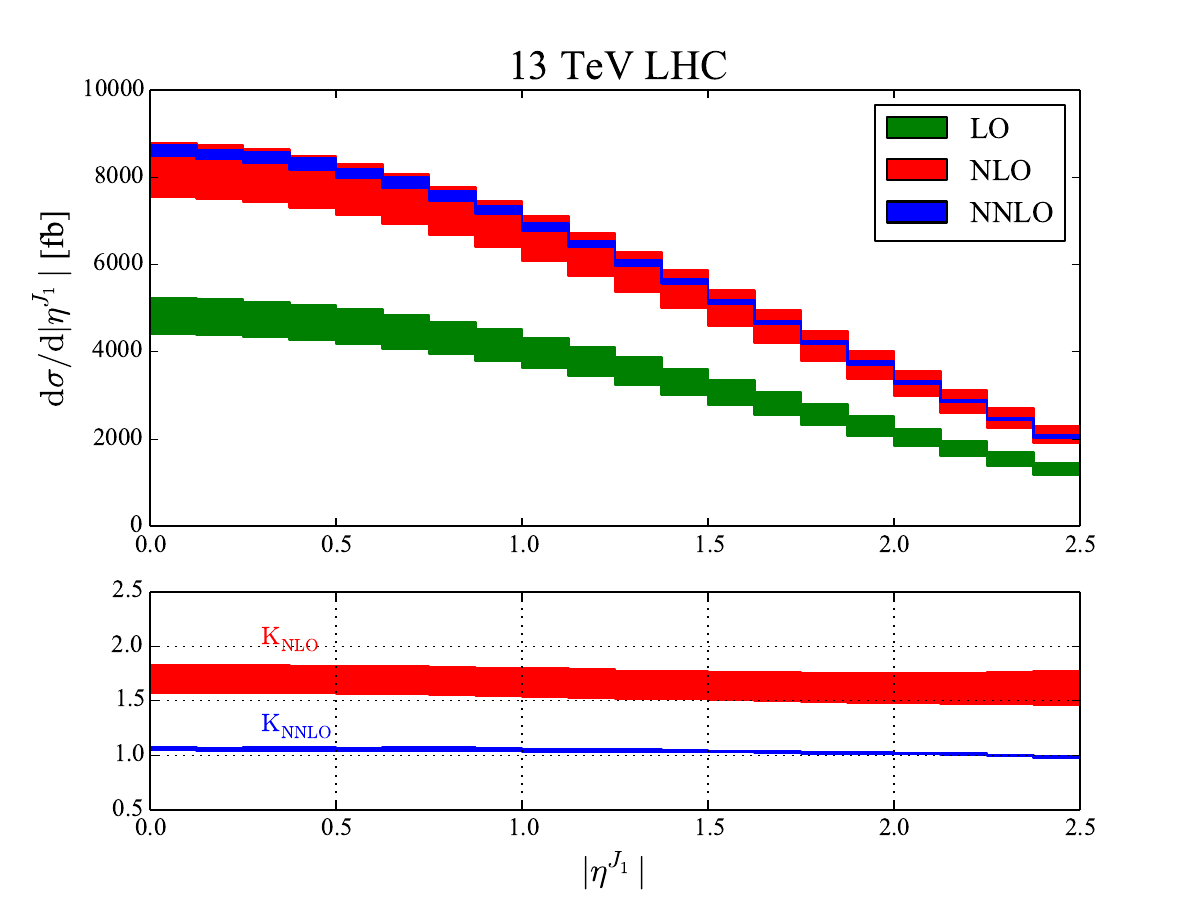} 
    \caption{Plots of the pseudorapidity distribution of the leading jet for both 8 TeV and 13 TeV collisions.  In each plot the upper inset shows the LO, NLO and NNLO distributions, while the lower inset shows $K_{\NLO}$ and $K_{\NNLO}$.  The bands indicate the scale variation.}
    \label{fig:etaJ1}
\end{figure}

The pseudorapidity distribution of the leading jet is shown in Fig.~\ref{fig:etaJ1}.  Both the NLO and NNLO $K$-factors are flat over the range of $|\eta^{J_1}|$ considered in the analysis.  The residual scale dependence at NNLO remains at the $1-2\%$ level.   The results for 8 TeV and 13 TeV collisions are similar.
 
\begin{figure}[h]
    \includegraphics[width=.49\linewidth]{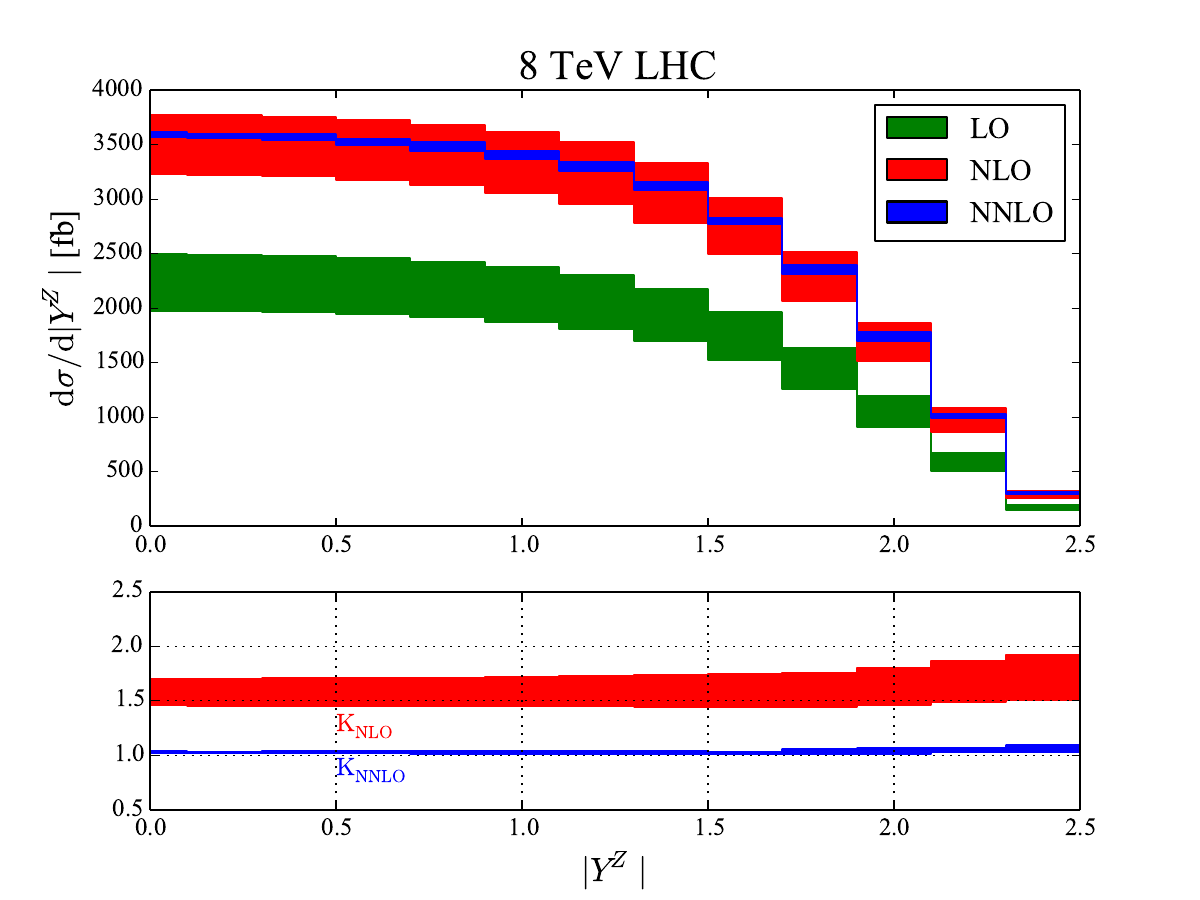} 
    \includegraphics[width=.49\linewidth]{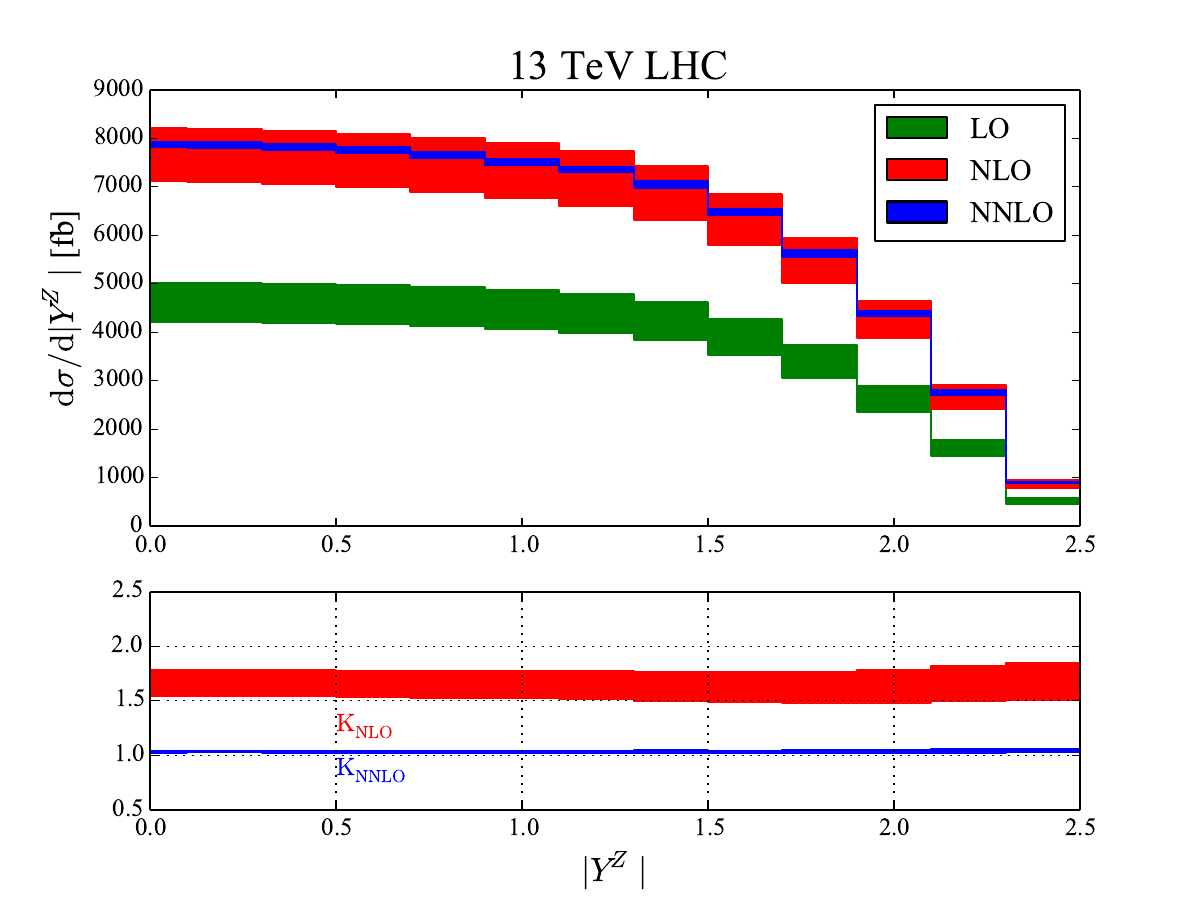} 
    \caption{Plots of the $Z$-boson rapidity distribution for both 8 TeV and 13 TeV collisions.  In each plot the upper inset shows the LO, NLO and NNLO distributions, while the lower inset shows $K_{\NLO}$ and $K_{\NNLO}$.  The bands indicate the scale variation.}
    \label{fig:Zrap}
\end{figure}

In Fig.~\ref{fig:Zrap} we show the rapidity distribution of the dilepton pair, which we label as the reconstructed $Z$-boson.   Both the NLO and NNLO corrections are nearly completely flat as a function of $Y^Z$. Both the magnitudes of the corrections, and the theoretical error as estimated by scale variation, are the same as for the fiducial cross sections shown in Table~\ref{tab:fidinc}.
 
\begin{figure}[h]
    \includegraphics[width=.49\linewidth]{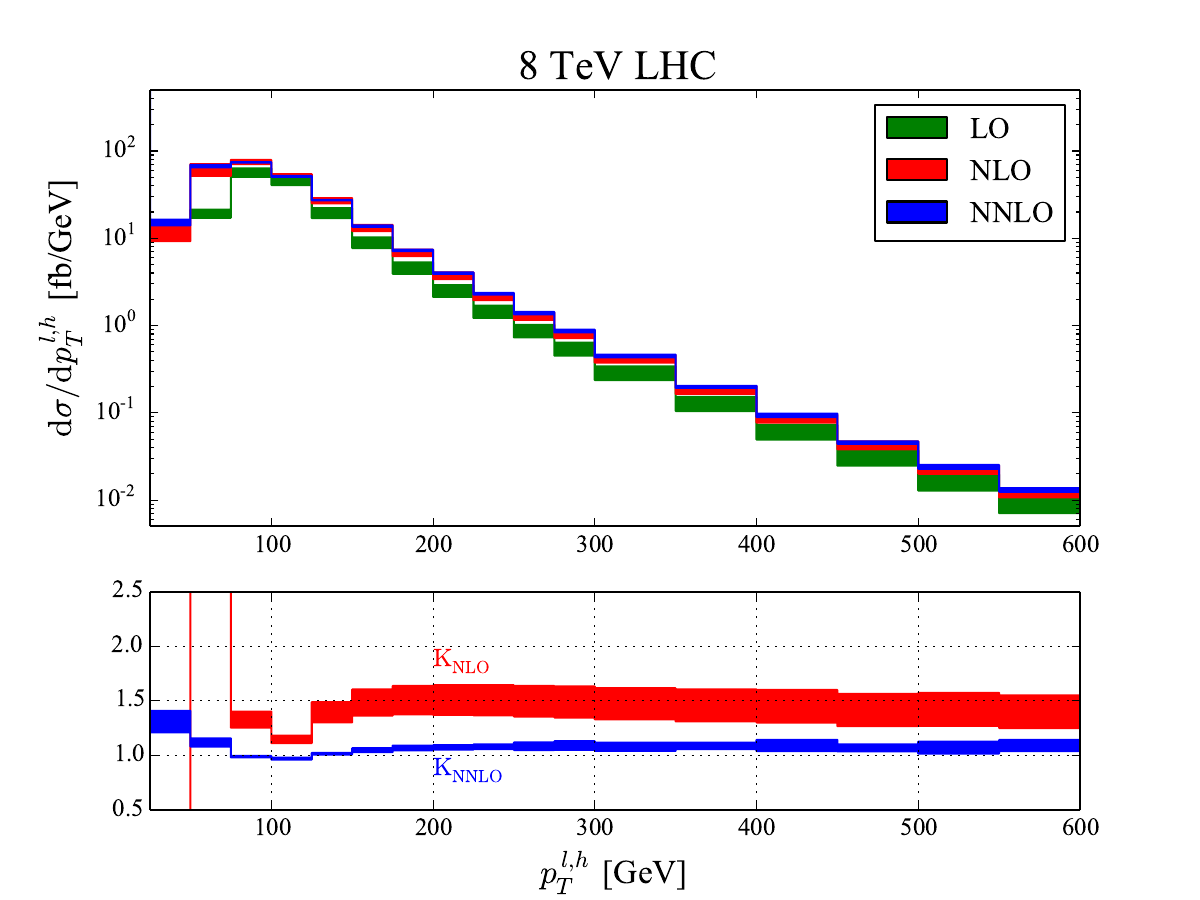} 
    \includegraphics[width=.49\linewidth]{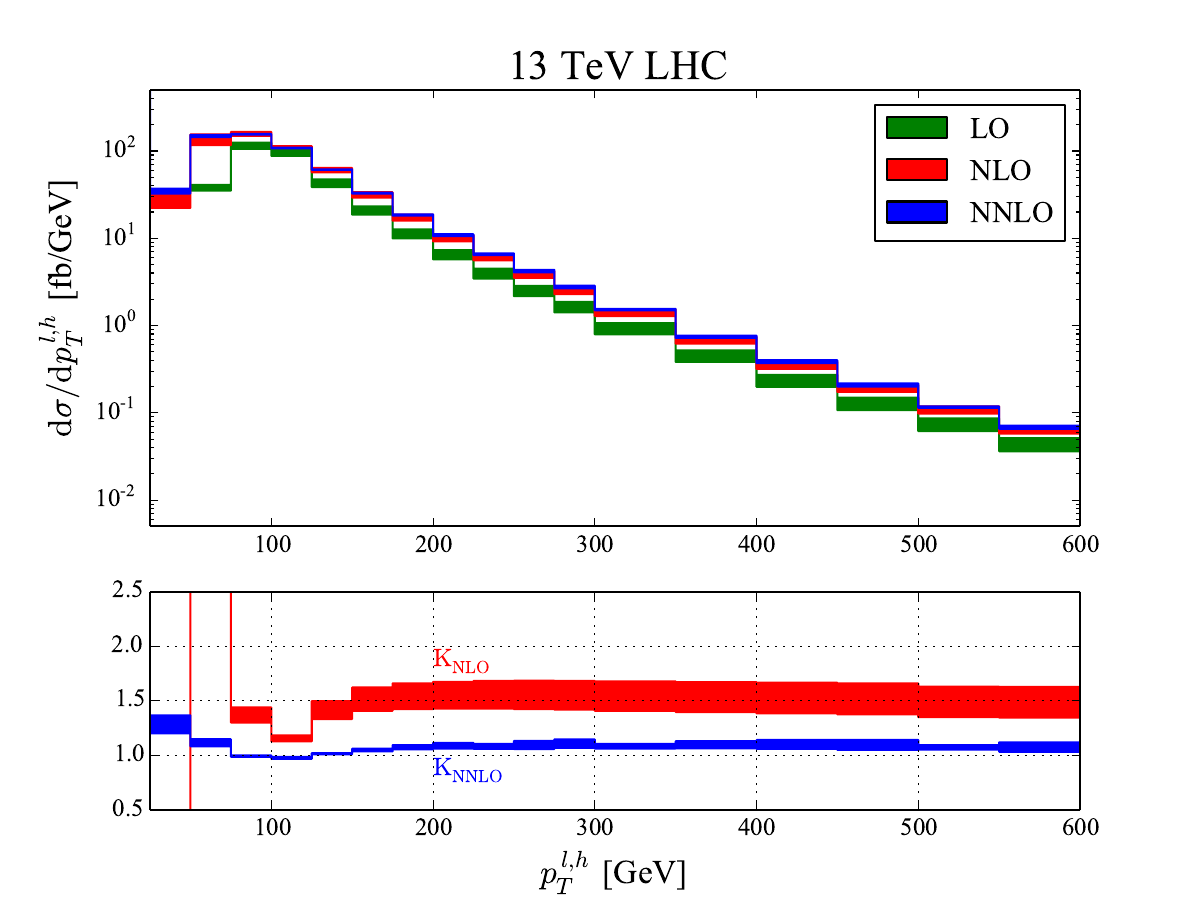} 
    \caption{Plots of the transverse momentum distribution of the leading lepton for both 8 TeV and 13 TeV collisions.  In each plot the upper inset shows the LO, NLO and NNLO distributions, while the lower inset shows $K_{\NLO}$ and $K_{\NNLO}$.  The bands indicate the scale variation.}
    \label{fig:pTlh}
\end{figure}

We now proceed to study the distributions of the leptons coming from the decay of the $Z$-boson.  We order the leptons in transverse momentum, and begin by studying the harder one, which we label with the superscript $h$.  The transverse momentum distribution of the leading lepton is shown in Fig.~\ref{fig:pTlh}.   To explain the observed pattern of corrections, we recall that the transverse momentum of the $Z$-boson is restricted to $p_T^Z>100$ GeV at LO due to the cut on the leading jet.  Since the leptonic decay products of the $Z$-boson must inherit this momentum, the hardest lepton must have $p_T^{l,h}>50$ GeV.  At NLO, when the restriction is relaxed, the leading lepton can have a smaller transverse momentum.  Near the LO boundary the cross section is sensitive to soft gluon effects, leading to the observed large NLO correction near $p_T^{l,h}=50$ GeV.  For higher values of $p_T^{l,h}$ the NNLO correction rises to a maximum near 15\% for both collisions energies.  The NNLO scale dependence is reduced to the few-percent level for $p_T^{l,h}>100$ GeV.
 
\begin{figure}[h]
    \includegraphics[width=.49\linewidth]{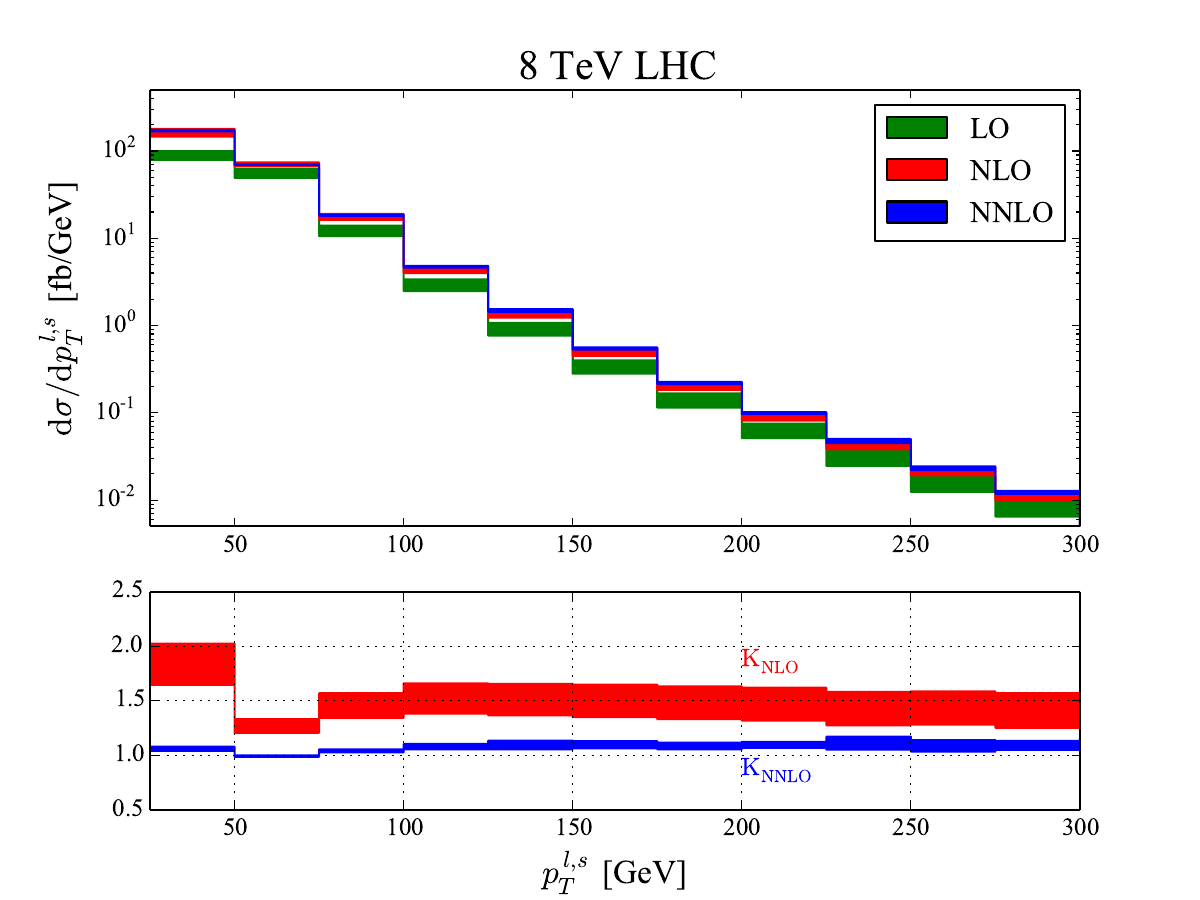} 
    \includegraphics[width=.49\linewidth]{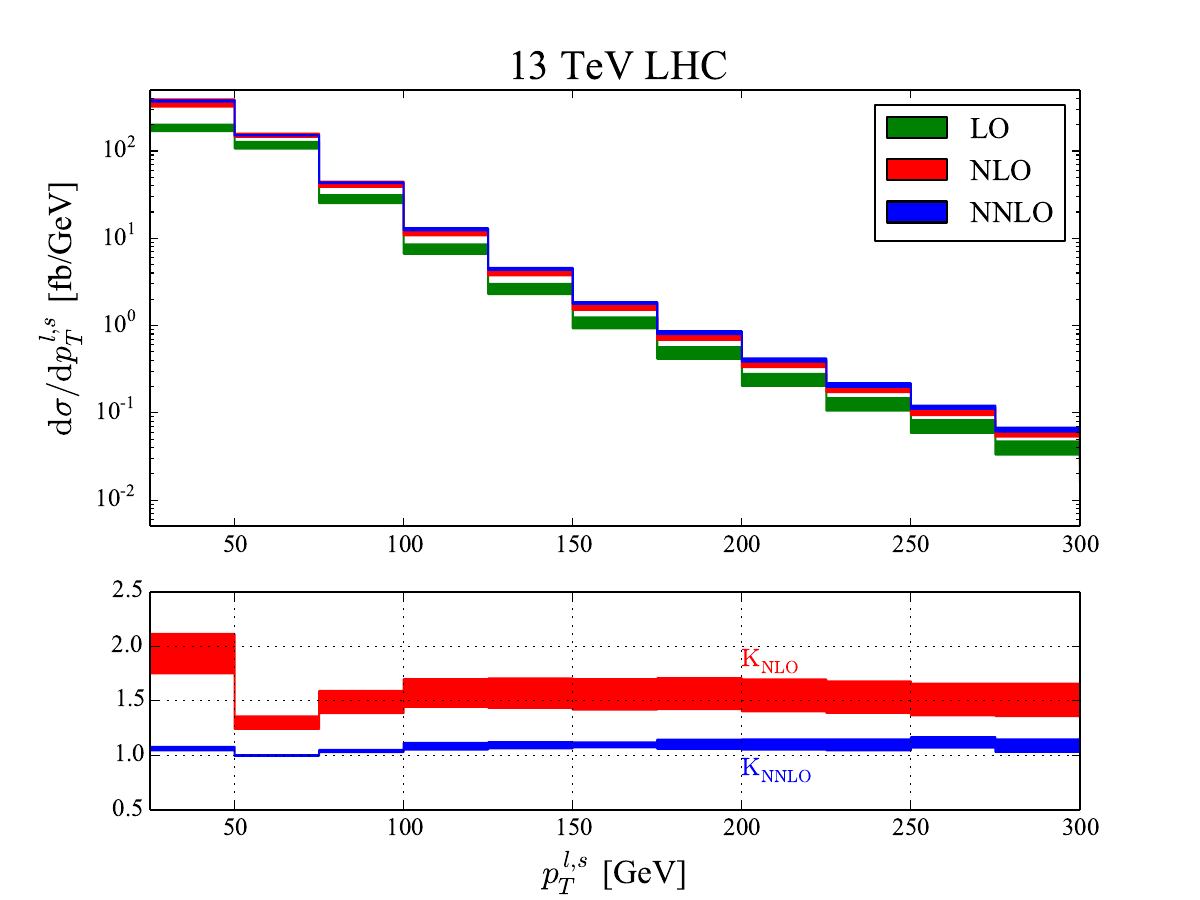} 
    \caption{Plots of the transverse momentum distribution of the softer lepton for both 8 TeV and 13 TeV collisions.  In each plot the upper inset shows the LO, NLO and NNLO distributions, while the lower inset shows $K_{\NLO}$ and $K_{\NNLO}$.  The bands indicate the scale variation.}
    \label{fig:pTls}
\end{figure}

The transverse momentum distribution of the softer lepton is shown in Fig.~\ref{fig:pTls}.  In this case there is no LO kinematic boundary as there is for the harder lepton. However, the opening up of the phase space region for $p_T^Z<100$ GeV at NLO does lead to a larger NLO $K$-factor at low $p_T^{l,s}$, since such leptons are preferentially emitted by low-$p_T^Z$ gauge bosons.  The NNLO corrections to the high-$p_T^{l,s}$ region again reach a maximum of 15\% for the highest values shown on the plots.  The scale variation is reduced to the few-percent level at NNLO.
 
\section{Comparison of predictions for different PDFs}
\label{sec:PDFs} 
 
Since the estimated theoretical error for $Z$+jet production from uncalculated QCD corrections is at the percent level over most of phase space upon inclusion of the NNLO corrections, it is interesting to consider what can now be learned from experimental measurements of this process.  It was pointed out that the $Z$-boson transverse momentum distribution can be used to discriminate among different models for the high-$x$ gluon PDF~\cite{Malik:2013kba}.  At the time the $Z$+jet process was only known at NLO, and theoretical errors masked the discrepancies between different PDF sets.  We revisit this suggestion now that the NNLO prediction is available.  In addition to the previously considered CT14 PDFs, we study the following alternative extractions:  NNPDF3.0~\cite{Ball:2014uwa}, MMHT2014~\cite{Harland-Lang:2014zoa}, and ABM12~\cite{Alekhin:2013nda}.  We focus on 13 TeV collisions in this section.
 
We begin by considering the fiducial cross section predicted by each of these PDF sets, assuming the cuts of Eq.~(\ref{eq:cuts1}).  The NNLO cross sections are shown in Table~\ref{tab:fidPDF}.  We recall the residual scale variation of the CT14 prediction from Table~\ref{tab:fidinc}: $14.87^{+0.05}_{-0.28} $ pb.  The MMHT2014 and NNPDF3.0 predictions are within this range, while the ABM12 result is not.  This indicates that the fiducial cross section does have sensitivity to the various PDF extractions; given precise experimental data, these different cross section predictions can be tested.  We note that this is not feasible without the NNLO corrections.  The residual scale uncertainty at NLO is nearly 10\%, which is larger than the observed differences.  We note that the fiducial cross section is dominated by Bjorken-$x$ values $x \sim \text{few} \times 10^{-2}$.  This is near the relevant region for Higgs boson production, indicating that $Z$+jet may help resolve PDF differences in this important region.
 
\begin{table}[h]
\begin{tabular}{|c|c|c|c|}
\hline
 CT14 & ABM12 & MMHT2014 & NNPDF3.0 \\
 \hline\hline
14.87 pb & 14.33 pb & 14.95 pb & 14.77 pb \\
\hline
\end{tabular}
\caption{Fiducial cross sections assuming different PDFs for the $Z$+jet process for 13 TeV collisions, using the cuts of Eq.~(\ref{eq:cuts1}). }
\label{tab:fidPDF}
\end{table}  
 
\begin{figure}[h]
    \includegraphics[width=.6\linewidth]{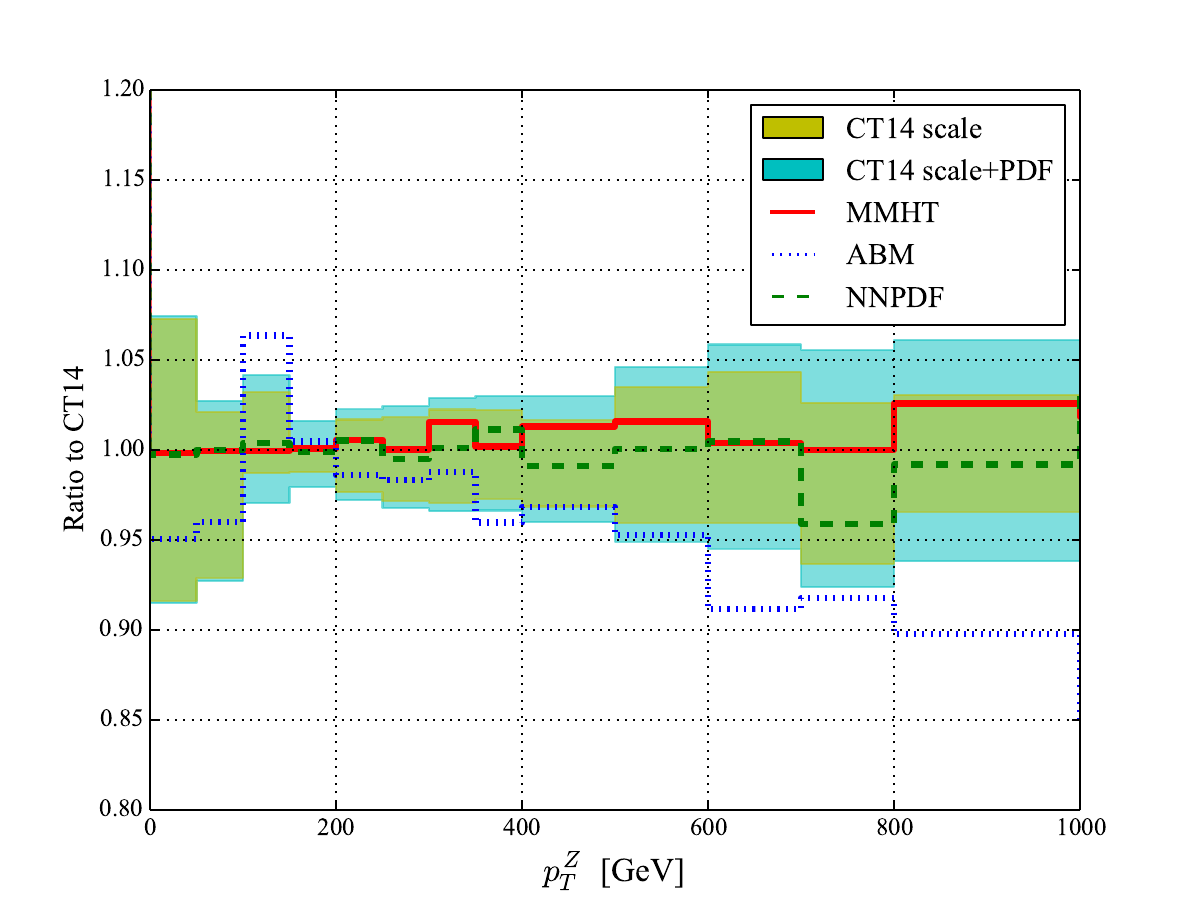} 
    \caption{The prediction of several PDF sets for the $Z$-boson transverse momentum distribution, shown as a ratio to the CT14 result.  The distribution for each PDF choice has been normalized by the fiducial cross section for that set.  The bands indicates the scale variation of the CT14 result, and the combined PDF+scale uncertainty.}
    \label{fig:pTZPDF}
\end{figure}

We next study the sensitivity of various differential distributions to the choice of the PDF.  To remove the differences in overall normalization already discussed above, we divide the distribution for each PDF choice by the fiducial cross section for that set in order to focus on shape differences.  We  consider three observables: $p_T^Z$, $p_T^{J_1}$, and $H_T$.  We have checked that other observables such as the $Z$-boson rapidity and the pseudorapidity of the leading jet have little sensitivity to the differences in PDFs.  The transverse momentum distributions of the $Z$-boson for the various PDF choices are shown in Fig.~\ref{fig:pTZPDF}.  For the CT14 set we show two bands indicating the theoretical scale error, and the combined PDF plus scale error obtained by combining these two sources of uncertainty in quadrature.  The scale dependence is at the few percent level in the high-$p_T^Z$ region, while the combined PDF+scale error is slightly larger. Both the NNPDF and MMHT distributions agree well with the CT14 result, and lie within the error bands.  The ABM result exhibits a different shape as a function of $p_T^Z$.  For $p_T^Z>400$ GeV it lies outside of the CT14 uncertainty band, and deviates from the other predictions by as much as 15\% at high $p_T^Z$.  A measurement of the high-$p_T^Z$ region can distinguish between the various PDF extractions.
 
\begin{figure}[h]
    \includegraphics[width=.6\linewidth]{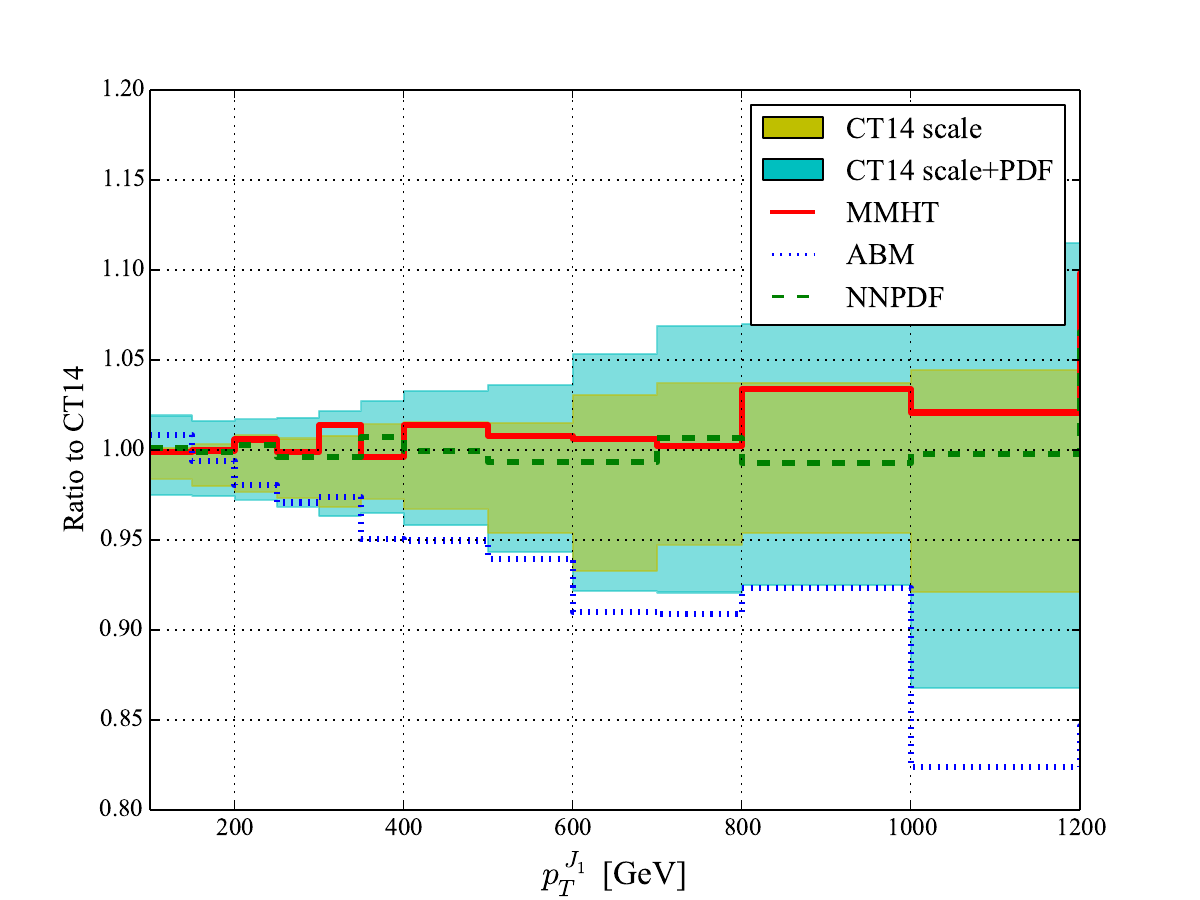} 
    \caption{The prediction of several PDF sets for the leading-jet transverse momentum distribution, shown as a ratio to the CT14 result.  The distribution for each PDF choice has been normalized by the fiducial cross section for that set.  The bands indicates the scale variation of the CT14 result, and the combined PDF+scale uncertainty.}
    \label{fig:pTJ1PDF}
\end{figure}

We show the comparison of the leading-jet transverse momentum distribution for the four PDF sets in Fig.~\ref{fig:pTJ1PDF}.  The scale dependence of the prediction grows as $p_T^{J_1}$ is increased, reaching the 10\% level in the TeV region.  Both the NNPDF and MMHT distributions lie within the uncertainty bands.  The ABM prediction differs from the other sets.  For $p_T^{J_1}>300$ GeV it lies outside the theoretical error band, indicating that this distribution can also be used to distinguish between the various PDF sets.
 
\begin{figure}[h]
    \includegraphics[width=.6\linewidth]{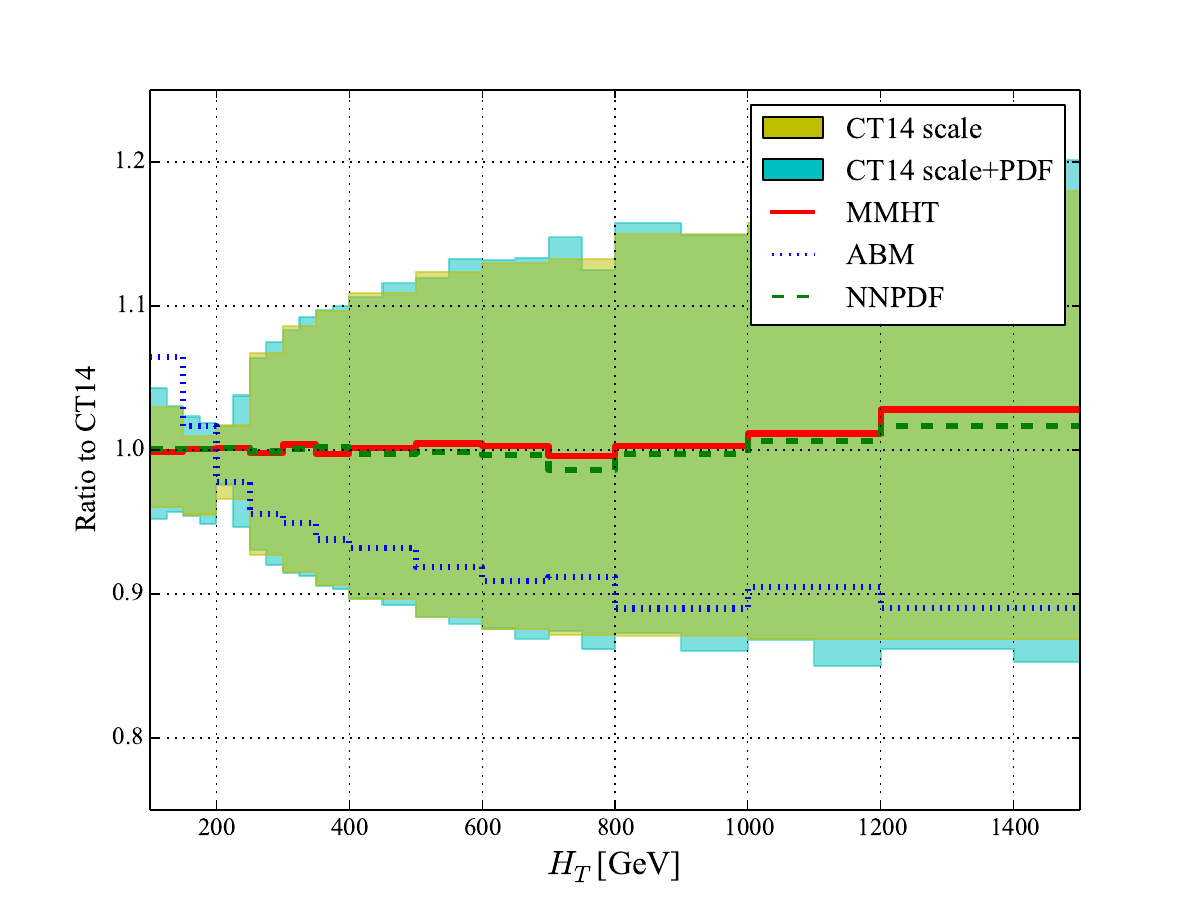} 
    \caption{The prediction of several PDF sets for the $H_T$ distribution, shown as a ratio to the CT14 result.  The distribution for each PDF choice has been normalized by the fiducial cross section for that set.  The bands indicates the scale variation of the CT14 result, and the combined PDF+scale uncertainty.}
    \label{fig:HTPDF}
\end{figure}

Finally, we show in Fig.~\ref{fig:HTPDF} the $H_T$ distribution for the various PDF sets.  The trends observed for each distribution are similar to those found for the transverse momentum distributions.  Both the MMHT and NNPDF sets agree well with CT14.  The ABM prediction exhibits a different shape; at low $H_T$ it is higher than the CT14 result, and it becomes up to 10\% lower at high $H_T$.  However, in this case the residual uncertainty is large and grows with $H_T$.  The differences between the distributions are masked by the theory error on the prediction.

\section{Summary and conclusions} 
\label{sec:conc} 

In this paper we have performed a detailed phenomenological study of the NNLO QCD corrections to the $Z$+jet process.  We have considered a large variety of distributions for both 8 TeV and 13 TeV collisions.  The NNLO corrections are in general small, increasing the cross section by a few percent.  The corrections grow larger in the high transverse momentum region, reaching the $15-20\%$ level in the TeV range for the $p_T^{J_1}$ distribution.  The high $H_T$ region is increased by a factor of two at NNLO.  The residual scale uncertainty of the NNLO prediction is at the few percent level, except in the high-$p_T^{J_1}$ and high-$H_T$ tails, where it reaches $15-20\%$.
 
Motivated by the excellent convergence and small residual uncertainty of the $Z$+jet perturbative expansion, we have studied the suggestion that high transverse momentum $Z$+jet production can be used to discriminate between different PDF sets~\cite{Malik:2013kba}.  We find that several observables are sensitive to different PDF choices: the fiducial cross section, and the shapes of the high $p_T^Z$ and $p_T^{J_1}$ distributions.  This sensitivity only arises at NNLO; at NLO, the residual scale uncertainty masks the differences between the various PDFs.

The work presented here is an important step toward preparing theoretical predictions to confront the upcoming data from LHC Run II.  Future work should include the combination of NNLO QCD with electroweak corrections, and the resummation of large logarithms that appear for certain observables such as the exclusive one-jet cross section~\cite{Boughezal:2015oga}.  We look forward to these future developments.
 
\section*{Acknowledgments}
We thank T.~LeCompte for helpful discussions.  R.~B. is supported by the DOE contract DE-AC02-06CH11357.  X.~L. is supported by the DOE grant DE-FG02-93ER-40762.  F.~P. is supported by the DOE grants DE-FG02-91ER40684 and DE-AC02-06CH11357.  This research used resources of the National Energy Research Scientific Computing Center, a DOE Office of Science User Facility supported by the Office of Science of the U.S. Department of Energy under Contract No. DE-AC02-05CH11231.  It also used resources of the Argonne Leadership Computing Facility, which is a DOE Office of Science User Facility supported under Contract DE-AC02-06CH11357.

\end{document}